\documentclass{article}
\usepackage{subfigure,graphicx,fullpage,times,amsmath,commath,fancyhdr,multirow,booktabs}\usepackage{natbib,authblk,url}
{}
{}
{}
\newcommand{\vknew}[1]{#1}

\newcommand{\flow}{q}
\newcommand{\dens}{k}
\newcommand{\spd}{v}
\newcommand{\vreal}{\spd_\textrm{real}}
\newcommand{\vest}{\widetilde{\spd}}
\newcommand{\spdm}{\spd_m}
\newcommand{\vmax}{\spd_\textrm{max}}
\newcommand{\edieacc}{\accumulation_\textrm{Edie}}
\newcommand{\totalacc}{N}
\newcommand{\edieproduction}{\production_\textrm{Edie}}

\newcommand{\accfactor}{\lambda}

\newcommand{\netprodi}{\production_{\textrm{net},i}}
\newcommand{\netacci}{\accumulation_{\textrm{net},i}}
\newcommand{\estproduction}{\widetilde{\production}}
\newcommand{\nrobs}{n}
\newcommand{\RMSE}{\textrm{RMSE}}
\newcommand{\production}{P}
\newcommand{\accumulation}{A}
\newcommand{\rl}{L}
\newcommand{\aggtime}{T}
\newcommand{\penrate}{\eta_\textrm{traj}}

\newcommand{\Fig}{Figure~}
\newcommand{\fig}{figure~}
\newcommand{\eq}{equation~}
\newcommand{\tab}{table~}

\newcommand{\halvepagina}{.5\textwidth}

\newcommand{\citeh}{\citep}

\newcommand{\citev}[1]{\citeauthor{#1} \cite{#1}}

\begin{document}

\title{Estimating the urban traffic state\\ with limited traffic data using the MFD}

\author[1,*]{Victor L. Knoop}
\author[2]{Marianthi Mermygka}
\author[1]{J.W.C. van Lint}
\affil[1]{Delft University of Technology}
\affil[2]{Aimsun London}
\affil[*]{Corresponding author, {v.l.knoop@tudelft.nl}}
%
\maketitle
\section*{abstract}
Urbanization leads to an increase of traffic in cities. The Macroscopic Fundamental Diagram (MFD) suggests to describe urban traffic at a zonal level, in order to measure and control traffic. However, for a proper estimation, all data needs to be available. The main question discussed in this paper is: \textit{How to derive a network-wide traffic state estimate?} We follow up on literature suggesting to base the operational estimate on the speed of limited sample cars sharing floating car data (FCD). We propose an initial step by constructing an MFD based on FCD, which is then used in step 2, the operational traffic state estimation. For operational traffic state estimation, i.e., the real-time traffic state estimation, the penetration rate is unknown. For both steps, we assess the impact of errors in the estimation. In light of the errors, we also formulate an indicator which shows when the method would yield insensible results, for instance in case of an incident. 
The method has been tested using microsimulation. 
A 26\% error in the estimated average density is found for a FCD penetration rate of 1\%; increasing the penetration rate to 30\% reduces the error in estimated average density to 7\%.

\section{Introduction}

With the Macroscopic Fundamental Diagram (MFD) traffic operations can be described at a zonal level, i.e. at the level of a (sub)network. The MFD shows to which extent an increase in average vehicle density (proportional to the vehicle accumulation in an area) reduces average vehicle speed, and thereby possibly vehicle flow. Due to the high number of roads in a town and the many connections and turnings of vehicles, this higher aggregation level makes the MFD a very suitable tool to describe traffic operations in urban environments for many different applications. 

Like many traffic variables of interest, the variables that describe the MFD relationship, i.e. average density (or vehicle accumulation), network average speed and/or average flow (production), are difficult to observe directly and completely. In the ultimate and ideal case, trajectories of all vehicles in the network over a long time period, are available, and the MFD quantities can be derived directly using Edies' relationships \cite{Edi:1965}. Unfortunately, this situation will not likely occur any time soon, implying that we need to estimate the MFD variables from whatever data \textit{are} available. Currently, many cities have loop detectors installed on major roads and at approaches of major intersections. Moreover, it is expected that in the near future, data from some vehicles will be available. One might argue that this is already the case with some providers collecting this type of data. Examples are Google, Apple, TomTom, but also some vehicles with connecting capabilities (e.g., Tesla) will be able to send information on their position and speed to a centralized server. 

\vknew{\citev{gayah2013} propose a clever way to use traffic from FCD to obtain a traffic state, which can be used to support network-wide traffic control strategies. In order to test their methodology, they use a simulated network of the city of Orlando. Their methodology takes advantage of the fact that each point on the MFD is related to a unique speed value. Thus, knowing the average speed from the probe vehicles allows them to infer on the MFD to what density that speed corresponds. This is a very promising method, but requires the MFD to be known. \citev{gayah2013} use full information to get this MFD on beforehand, which is unrealistic to get.} 

In this paper we elaborate and expand on this method, assessing the uncertainties which arise when the method is being applied in practice. 
Three issues stand out. Firstly, regarding the estimation of the traffic state in normal conditions. An important step is that an MFD should be obtained a priori. We present a method to obtain an MFD using FCD and combine this with loop detector data. This yields errors, because of the low penetration rate of FCD and because of the uneven distribution of these vehicles sharing their data. Secondly, the observed speed of the equipped vehicles is not necessarily the same as the speed of all vehicles, which yields an estimation error. We will assess this error and the combination with the error in estimating the MFD. Thirdly, the MFD might change in incident conditions, which could lead to an estimated traffic state which is far from reality. We discuss how these situations can be assessed with the FCD at hand. The questions discussed in this paper are hence: \textit{How can we reliably estimate the network traffic state (average network density)}, and  
 \textit{How can  the (un)reliability of the estimate be indicated}?

The key methods laid out in the paper do exactly this. Roughly speaking, the traffic state method has two major steps. In the first step (initialization), FCD is fused with loop detector data to obtain an MFD. In the second step (operational estimates), the FCD speed is combined with the estimated MFD curve to estimate the traffic state (i.e,. the average density); this second step is not new, and follows the ideas of \citev{gayah2013}. The intermediate step of the MFD -- rather than a direct up-scaling of the number of equipped cars towards density -- means the penetration rate of vehicles sharing FCD does not need to be known in the second step, and moreover, the estimated traffic state is less dependent on fluctuations of the penetration rate. 
The second contribution is the estimation of the reliability of the estimated state. To this end, we consider the rate of change of speed of the FCD as indicator for a possible disruptive incident, which causes the MFD to change. Note that this paper summarizes the main results extensively elaborated in the thesis of \citev{Mer:2016}. 

The paper is organized as follows. In section \ref{sec_existing_methods} we briefly overview the literature related to state estimation in general and MFD estimation in particular. The method then is described in section \ref{sec_method}. We test it using microscopic simulation in section \ref{sec_simulations}. In this test, we vary the penetration rate of FCD from 1\% to 30\% to assess its capabilities under low penetration rates. We explicitly discuss the errors in the method, and the range of possible outcomes. An indicator for the reliability of the method is presented in section \ref{sec_reliability}; using simulation of incidents, this method is being tested. The paper closes with the conclusions (section \ref{sec_conclusions}). 

\section{Traffic estimation}\label{sec_existing_methods}

Estimating the traffic state using data from multiple sources is one of the key research themes within the traffic flow and ITS community today. Real-time traffic estimates provide the basis for traffic prediction; traffic management and control and for information provision to travelers and road authorities. Reliable traffic state estimates also result in consistent historical databases that can be used for modeling, analyses, evaluation of measures and policies, and for research itself. This section briefly introduces the challenges in estimating the traffic state in general (\ref{sec_TSE}); and motivates how the MFD can be used to do so on a network level (\ref{sec_litMFD}).

\subsection{Traffic state estimation}\label{sec_TSE}

As noted in the introduction, ideally, the traffic state at any level of scale (on a lane, link or entire network) is reconstructed by using trajectories of all vehicles. \citev{leclercq2014} shows that the trajectory method can produce the MFD very accurately in all network shapes and thus suggests that this method can be used as a base to evaluate and compare other methods---something feasible in simulation-based studies. The same advice holds for state estimation in general. To estimate state variables in the field we need to use estimation techniques in combination with whatever data \textit{are} available. Such traffic state estimation methods involve (apart from data from various types of sensors available): (1) theories (assumptions, models, equations) that describe the relation between the data and the state variables of interest; and (2) data assimilation techniques that combine data and model predictions and in the process address measurement and modeling errors. 

The most popular types of state estimation approaches are \textit{sequential Bayesian estimation techniques} that combine traffic flow simulation models with for example (extended) Kalman filters \cite{Hinsbergen2012a,Wang2005a}, ensemble Kalman filters \cite{Yuan2015a,Work2008a}, unscented Kalman filters \cite{Ngoduy2011a}, Particle filters \cite{Wang2016a}, and many variations on the same theme to assimilate data from a multitude of different sensors to estimate density; space mean speed; travel time; fundamental diagram parameters, and much more. The great appeal to these approaches, is that they combine knowledge of traffic dynamics with principled data assimilation techniques to "optimally" balance the uncertainty and limitations of traffic flow models with the inherent uncertainty in the data. The quotes indicate that in many cases (due to a variety of reasons) no optimality guarantees are possible. Nonetheless, there is a huge body of evidence that support the effectivity of Bayesian sequential estimation in traffic estimation (and in many other estimation problems). The great advantage is that these methods provide an integrated solution for network wide state estimation (and prediction), and that they use tractable behavioral and physical relationships, which make them highly suitable for what-if reasoning; control optimization and application under non-recurrent conditions. The price for this explanatory power is that these (simulation) model-based methods are generally complex to design and maintain, and sensitive to data errors, and in particular to systematic bias in these data. Moreover, they require many (partially unobservable) inputs (e.g. traffic demand, control inputs) and contain many parameters that need to be calibrated or even predicted from data. 
At the other end of the spectrum we find purely statistical (data-driven) methods, that use no (traffic flow) assumptions at all, but rely on statistical correlations to infer traffic variables on particular links over particular time periods from historical and real-time data. These approaches may be based on relatively simple linear models \cite{Kong2013a} or on highly advanced deep-learning architectures \cite{Yu2017a} and everything in between. These methods are becoming increasingly popular, particularly for large-scale network wide estimation and prediction, but they lack explanatory power to uncover unobserved state variables, like vehicular density.

Whereas an integrated Bayesian approach to estimate density in an entire (urban) network would  provide a powerful solution to estimate also network level quantities (by aggregation), the data requirements for such an approach seem infeasible in many cases still. For our purpose more parsimonious and direct approaches to estimate network density are preferable. One category of these approaches are what we term data-data-consistency approaches, that explicitely exploit semantical differences between different data (where one data source may provide supporting evidence to correct a second source), and utilise direct analytical or statistical relationships (e.g. travel time equals distance over speed, density equals the spatial derivative over cumulative flow, the fundamental relationship, etc) to infer state variables. Examples include methods that fuse counts with travel times to estimate vehicle accumulation \cite{Bhaskar2014a,VanLintHoogendoorn2015}; or methods that fuse local speeds and travel times to compute / correct space mean speeds \cite{Ou2008b}. In our case, we want to exploit the (shape of the) MFD itself, in the estimation of network density. To this end, we briefly explore different approaches to estimate the MFD (variables) from data.

\subsection{MFD estimation}\label{sec_litMFD}
If only detector data are available to obtain the MFD, the network flow can be calculated sufficiently and accurately. However, the same does not apply for the vehicle speed and the network density. As literature has shown \citeh{buisson2009,Cou:2011}, the location of the detectors can influence significantly the network speed estimation. When the detectors are near the stop line, most of the captured vehicle speeds are low and consequently, the situation further upstream of the traffic light is not taken into account. This issue seriously affects the validity of the detector data to estimate the MFD.

\citev{nagle2014} suggest a methodology that overcomes these disadvantages. Their method uses the generalized definitions of Edie \citep{Edi:1965} to estimate the network-wide variables from probe vehicle data. However, in order to apply these formulas, the data of all the vehicle trajectories are necessary, whereas usually only a small number of vehicles serves as probes. The authors suggest this difficulty can be overcome as long as the ratio of the probe vehicles is known. In order to acquire the ratio, they proposed dividing the number of vehicles that were tracked by  GPS in the analysis area for a specific time period to the number of vehicles that crossed the detectors in the same area and period. Microsimulation tests of their methodology showed that a 20\% probe penetration rate can provide accurate estimations for any traffic state, making this a robust methodology to acquire the MFD. Also this study has some limitations. For one, it assumes that the probe vehicles are uniformly distributed across the network, which is not realistic in many cases.

\citev{leclercq2014} also combines data from probe vehicles with loop detector data but in a slightly different way. From the probe vehicles, the average network speed is derived; and from the loop detectors, the average network flow. Their results show that a 20\% probe penetration rate can significantly improve the estimation of the network speed for the MFD. However, this study has the limitation that the dynamics of the network are not captured completely, since the loop detectors are not placed everywhere in the network. Also \citev{Amb:2016} describe a problem which is similar to ours: fusing loop detector data and FCD for MFD estimation. They succeed in finding a MFD with an elegant method. However, they need the loop detector data throughout the whole estimation process, i.e., for MFD estimation as well as for real-time traffic state estimation. 

\citev{du2015} tries to overcome these limitations  by proposing a MFD estimation method without the conditions that the probe penetration rate is homogeneous and the detectors are placed in all links. Their approach estimates the appropriate average probe penetration rates from the weighted harmonic means with the weights being the travel times and the distances of the  individual  probe vehicles. They test their methodology with microsimulation and the results show that it is a very effective approach. \citev{du2015} suggest that since mobile probe data are becoming more and more available, this methodology can soon be applied also with real data. 
\citev{tsubota2014}, finally, is one of the few studies that applied such a data fusion approach using real data for the city of Brisbane, Australia. Similarly to \cite{Bhaskar2014a,VanLintHoogendoorn2015}, their approach aims to estimate the network density combining loop detector data, traffic signal timings and probe vehicle data by exploiting the relationship between cumulative counts and travel times between consecutive detector locations. 

The common theme of these approaches is that data fusion seems indeed the way forward in estimating the MFD (variables). Unfortunately, without large-scale ground-truth data, we do need to resort to simulation to assess the quality of the methods. This is also our approach. We combine a number of elements of existing methods, and contribute to the existing literature by designing a data fusion MFD estimation method that works well for low penetration rates of FCD. Moreover, we develop a method that endogenously computes the quality of this estimate.

\section{Proposed traffic state estimator}\label{sec_method}

\begin{figure*}
	\centering
	\includegraphics[width=0.95\textwidth]{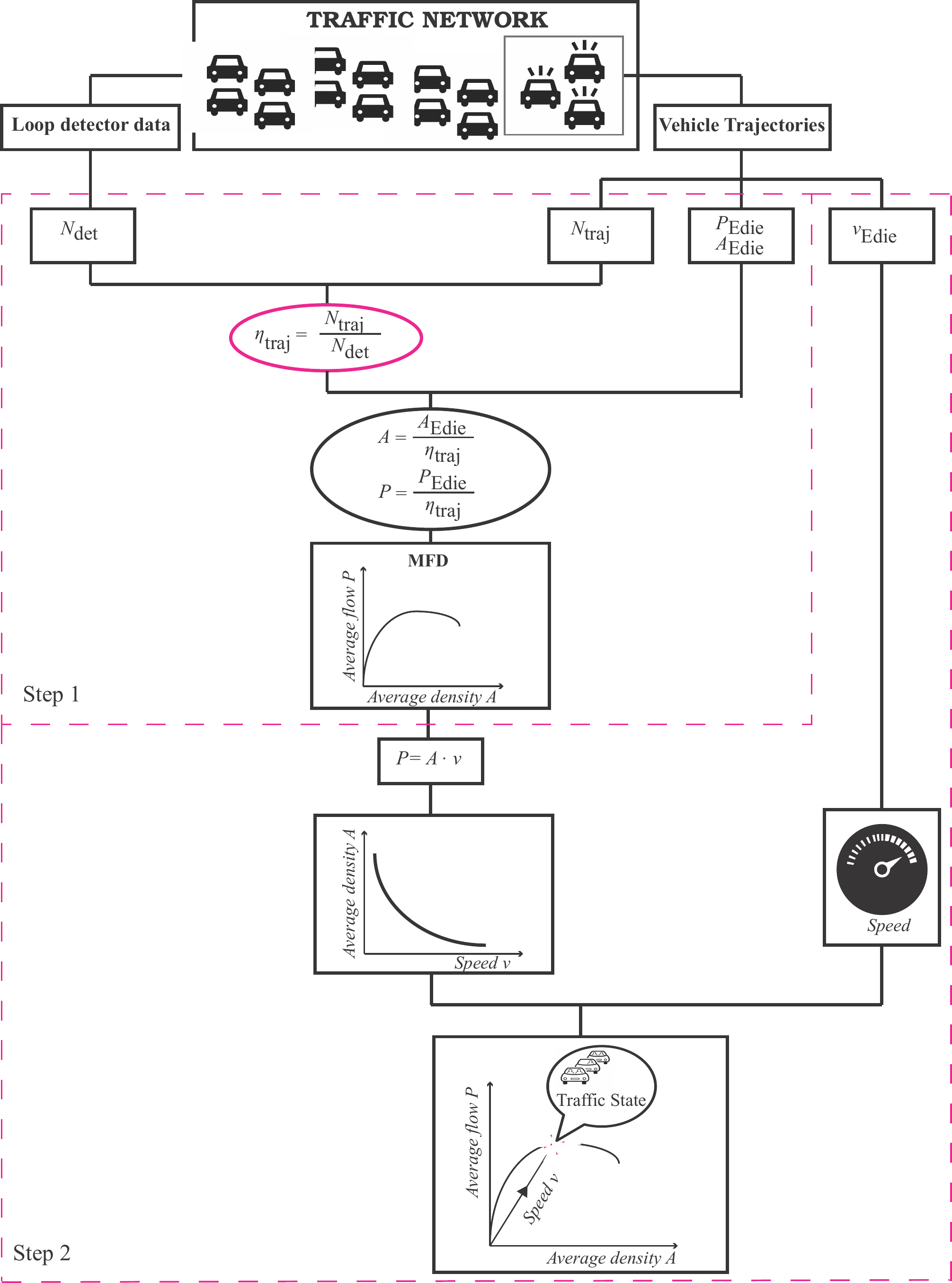}
	\caption{Schematic representation of the two steps of the traffic state estimation process}
	\label{mfd_fusion}
\end{figure*}

In this research, we use two stages in the estimation process. They are graphically shown in \fig \ref{mfd_fusion}. First, the MFD is estimated, and then with the use of the (speeds of the) FCD and the  MFD, the traffic state is found. 

\subsection{Step 1: creating MFDs}
The ground truth for the traffic operations is formed by the generalised speed, flow and density according to Edie's definition \citeh{Edi:1965}, scaled by the network length.
According to Edie's formulas, the flow $\flow$ in the network for any arbitrary area in space-time is calculated by dividing the total distance traveled for all vehicles over the amount of space (in this case, the total road link length in kms of lane $\rl$) and the amount of time (aggregation time, $\aggtime$). This hence also applies for the average flow, $\production$:
\begin{equation} \label{flow_ed} 
\edieproduction=\frac{\sum_{ i}^{n}d_{i}}{LT}
\end{equation}	
In this equation, $d_{i}$ is the travel distance of vehicle $i$ and $n$ is the total number of vehicles that use the network during the time interval. Note that the average flow is proportional to $\edieproduction$, differing by a scaling of network length.

Similarly, following Edie's definitions, the density $\dens$  is computed by dividing the total time spent over the amount of space  and amount of time. This also holds at the network scale, hence for the average density $\edieacc$ we find
\begin{equation} \label{density_ed} 
\edieacc=\frac{\sum_{ i}^{n}t_{i}}{LT}
\end{equation}
In this equation, $t_{i}$ is travel time of vehicle $i$. The total number of vehicles in the system, the accumulation $\totalacc$, can be derived from the average density:
\begin{equation} \label{accum_ed} 
\totalacc=\edieacc \rl
\end{equation}

The speed $\spd$ can be determined by the ratio of average flow and average density:
\begin{equation} \label{speed_ed} 
\spd=\frac{\flow}{\dens}=
\frac{\production}{\accumulation}=\frac{\sum_{i}^{n}d_{i}}{\sum_{i}^{n}t_{i}}	
\end{equation} 	

The first step in the method is to create the MFD. This is done by combining loop detector data and floating car data. The main idea on how to fuse the data is based on the fact that the vehicle trajectories offer representative information, but only a subset of them is available. However, if the relative size of the subset (the penetration rate) is known, then the variables calculated from the subset can be scaled to the full set of vehicles. In other words, if the penetration rate of the known vehicle trajectories  in relation to the total number of vehicles (denoted $\penrate$) is calculated, the information from the subset of vehicle trajectories can be divided with the subset to represent the entire network traffic state. 

This penetration rate of the known vehicle trajectories can be calculated by relating the total number of vehicles that cross the detectors in every time interval with the number of vehicle trajectories that traverse the links with the respective detectors. The average flow and the average density calculated by the subset of the known vehicle trajectories using Edie's definitions can then be divided with the proportion to represent the total set of vehicles.  In an algorithm format, the proposed data fusion approach of the 1\textsuperscript{st}  step is:
\begin{enumerate}
 	\item From the detector data, calculate the total number of vehicles crossing the detectors in the aggregation period $N_\textrm{det}$.
	\item For the subset of known vehicle trajectories, calculate the average density $\edieacc$ (Equation \ref{density_ed}), the average flow $\edieproduction$ (Equation \ref{flow_ed}), as well as the number of vehicle trajectories that traverse the location where detectors  are located, $N_\textrm{traj}$.	
	\item For each time interval, measure the total number of counts at all detectors combined ($N_\textrm{traj}$), and calculate which fraction of this number is represented in the known vehicle trajectories:	
	\begin{equation}
	\penrate=\frac{N_\textrm{traj}}{N_\textrm{det}}
	\end{equation} 
It is important to note that in the vehicle number, we simply added all observations for all detectors (i.e., we do not calculate so on a link-by-link basis), and use the law of large numbers which apparently also works in constructing a MFD in the first place. The found value $\penrate$ can be used as estimate for the penetration rate. 
	\item For each time interval, calculate the average density $\accumulation$ and average flow $\production$, as
	\begin{equation}
	\accumulation=\frac{\edieacc}{\penrate}
	\end{equation}	
	\begin{equation}
	\production=\frac{\edieproduction}{\penrate}
	\end{equation}		
	\item Construct the MFD of the network, finding the mathematical function that best fits the data points of $\accumulation$ and $\production$.	
\end{enumerate}

\subsection{Fitting a functional form MFD}
Step 2 of the method requires a functional form of the MFD.
The requirements that the formula of the MFD  needs to fulfil are:

\begin{enumerate}\itemsep0em 
	\item The formula needs to follow a concave shape, due to the  alternating trend of the two MFD branches: the free flow and the congested branch.  In the free flow branch, the flow increases rapidly when the density increases and the slope of the curve is quite steep. In the congested branch, the flow decreases gradually with the density increase, so the slope is less steep. 
	
	\item The MFD has zero flow for zero density, so the fitting formula needs to have zero constant.
	
	\item The derivative of the MFD function at zero density should be equal to the free flow speed.
\end{enumerate}
The often used expression for a third order polynomial fit without a constant (e.g., \citeh{Kno:2013TRRGMFD_A10}) does not fulfil the first criterion; initial fits in fact show that the second turning point of the third order polynomial is often lower than jam density, hence the flow of the fitted fundamental diagram \emph{increases} for high densities. 

Therefore, we fit another function with not more parameters. We opt for a combination of a Greenshields \citeh{Gre:1934} fundamental diagram (parabolic) with elements of a Drake fundamental \citeh{Dra:1967} diagram (exponential decay). We fit
\begin{equation}
\production=p_1 \cdot \accumulation \cdot e^{\accfactor \accumulation} + p_2 \cdot \accumulation^2 \cdot e^{2 \accfactor \accumulation}
\end{equation}
Examples of the functional form can be found in section \ref{sec_MFDresults}. 

This function fulfils the criteria and at the same time has only 3 parameters to calibrate: $p_1$, $p_2$ and $\accfactor$. An interpretation for $\accfactor$ is the non-linearity in the density; with $\accfactor=0$ one would obtain a symmetrical parabola (Greenshields fundamental diagram). 

The parameters are found using the data points $\netprodi$ and $\netacci$. The subscript $i$ denotes an individual measurement point over a time interval (and a spacial extent). We estimate the matching average flow according to the average density at data point $i$, and the MFD:
\begin{equation}
\estproduction_i=
p_1 \cdot \netacci \cdot e^{\accfactor \netacci}
 + p_2 \cdot \netacci^2 \cdot e^{2 \accfactor \netacci}
\end{equation}
Next, we take the root mean squared error of average densities for all time intervals:
\begin{equation}
\RMSE=\sqrt{\left.\sum_{i=1}^{\nrobs}\left(\left(\estproduction_i-\netprodi\right)^2\right)\right/\nrobs}
\end{equation}
The RMSE depends on the parameters. Fitting the MFD consists of adjusting the parameters such that the RMSE is minimized. We use a \textit{fminsearch} function in Matlab for this.

Using the fundamental relationship of \eqref{speed_ed},  we can describe the speed $\spd$ as mathematical function of the average density $\accumulation$:
\begin{equation}
\spd(\accumulation)=\frac{\production}{\accumulation}=p_1 \cdot e^{\accfactor \accumulation} + p_2 \cdot \accumulation \cdot e^{2 \accfactor \accumulation}
\end{equation}
If the parameters are known, we also have the relationship between average density and speed.

\subsection{Step 2: using FCD and MFDs to find the traffic state}\label{sec_step2}
The MFD obtained from the 1\textsuperscript{st} step can be used in the 2\textsuperscript{nd} step to calculate the traffic state based on FCD  from an unknown penetration rate. The second part of the schematic flow chart in Figure \ref{mfd_fusion} depicts the 2\textsuperscript{nd} step of the process.

In the MFD relationship for the network, each speed value corresponds to a unique density value. Taking advantage of this, speed data from any data source can be used to derive the average network density and thus, the point on the MFD that the traffic network is. In an algorithm format, the proposed density estimation approach of the 2\textsuperscript{nd} step is:
\begin{enumerate}
\item Given the relationship between speed and density for the network and speed measurements from any available data source, calculate the average network density. 
\item Use the derived average network density to indicate the traffic state on the MFD.
\end{enumerate}

\section{Case study}\label{sec_simulations}

The process described in section \ref{sec_method} is tested using a case study. The setup of this test is described in section \ref{sec_expsetup} and the results in section \ref{sec_MFDresults}.
\subsection{Experimental setup}\label{sec_expsetup}
For the verification of the method we use microsimulation. That way, we have the ground truth data available in the form of trajectories, hence the ground truth average density and average flow can be determined using \eqref{accum_ed} and \eqref{flow_ed}. In this case, we will be using S-Paramics. 

\begin{figure}
\subfigure[Map (source: OpenStreetMap)]{\includegraphics[width=\halvepagina]{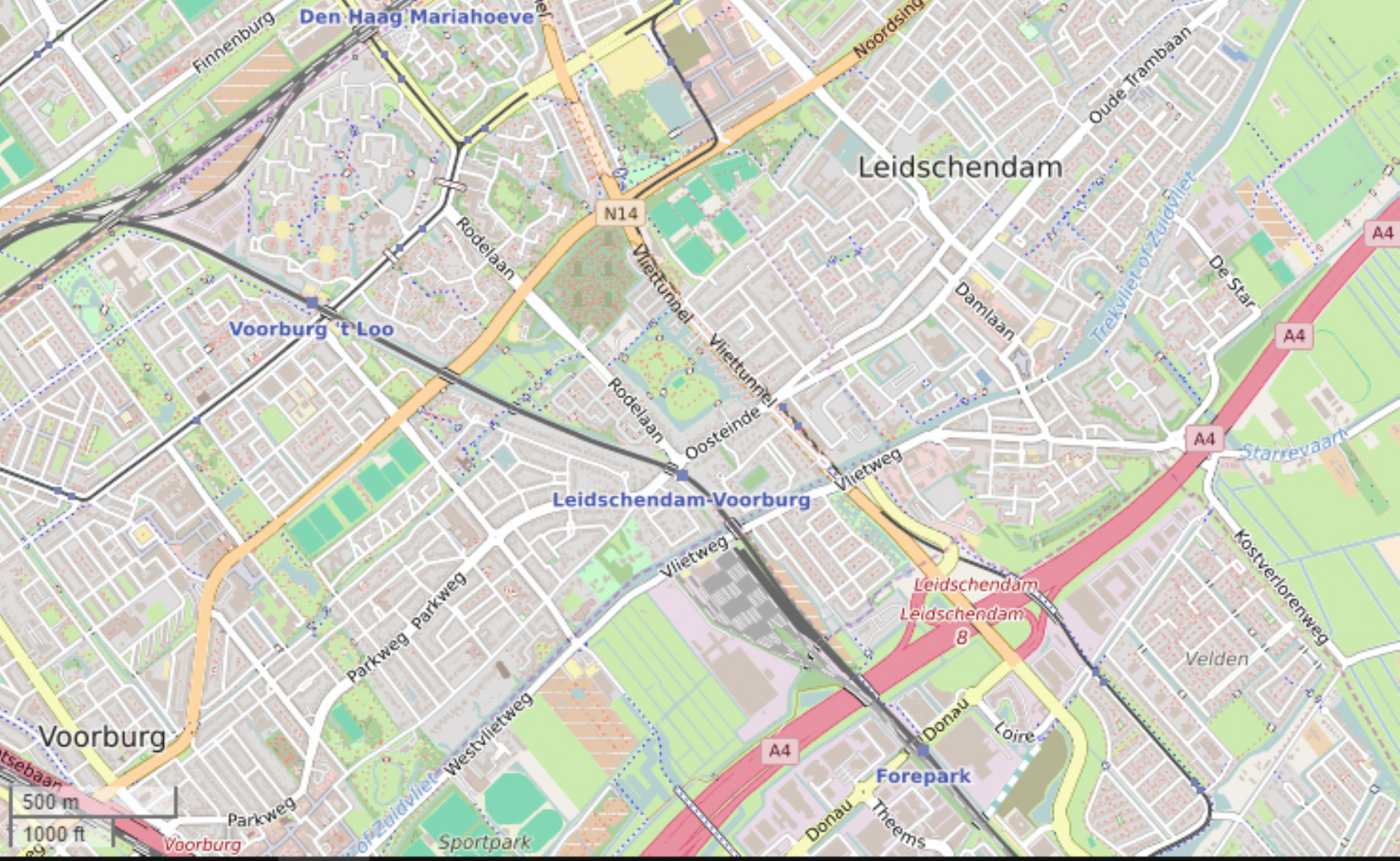}\label{map}}
\subfigure[The simulated network with the location of the loop detectors  indicated with green lines, and the location of the incidents (letters)]{\includegraphics[width=\halvepagina]{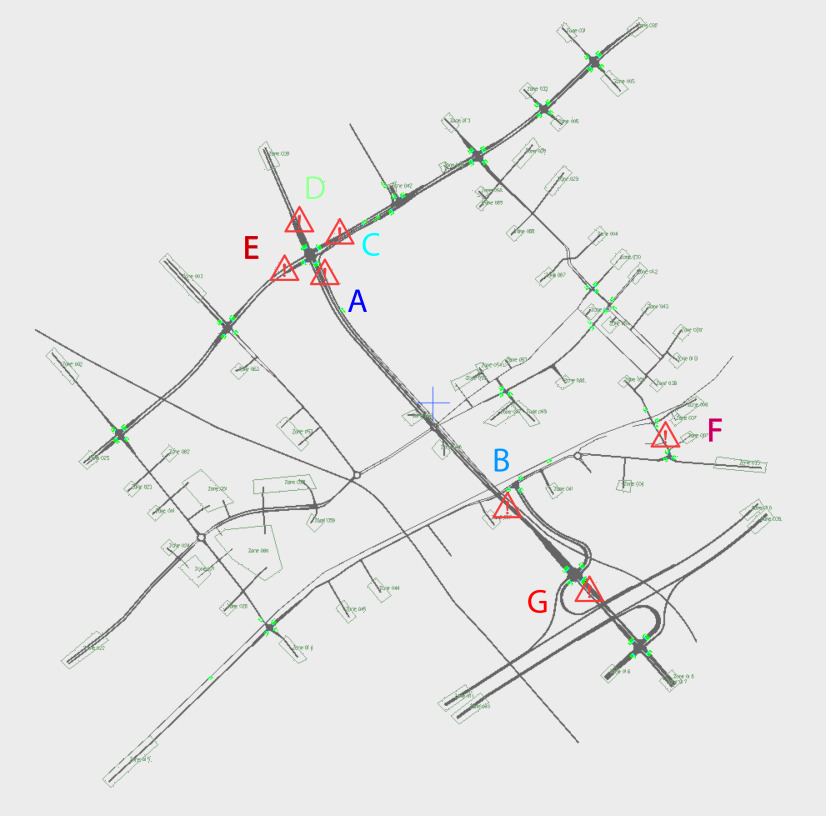}
\label{fig:leidschendam_model}\label{incident_locations}}
\subfigure[Traffic demand during the morning peak period of the  simulation]{\includegraphics[width=\halvepagina]{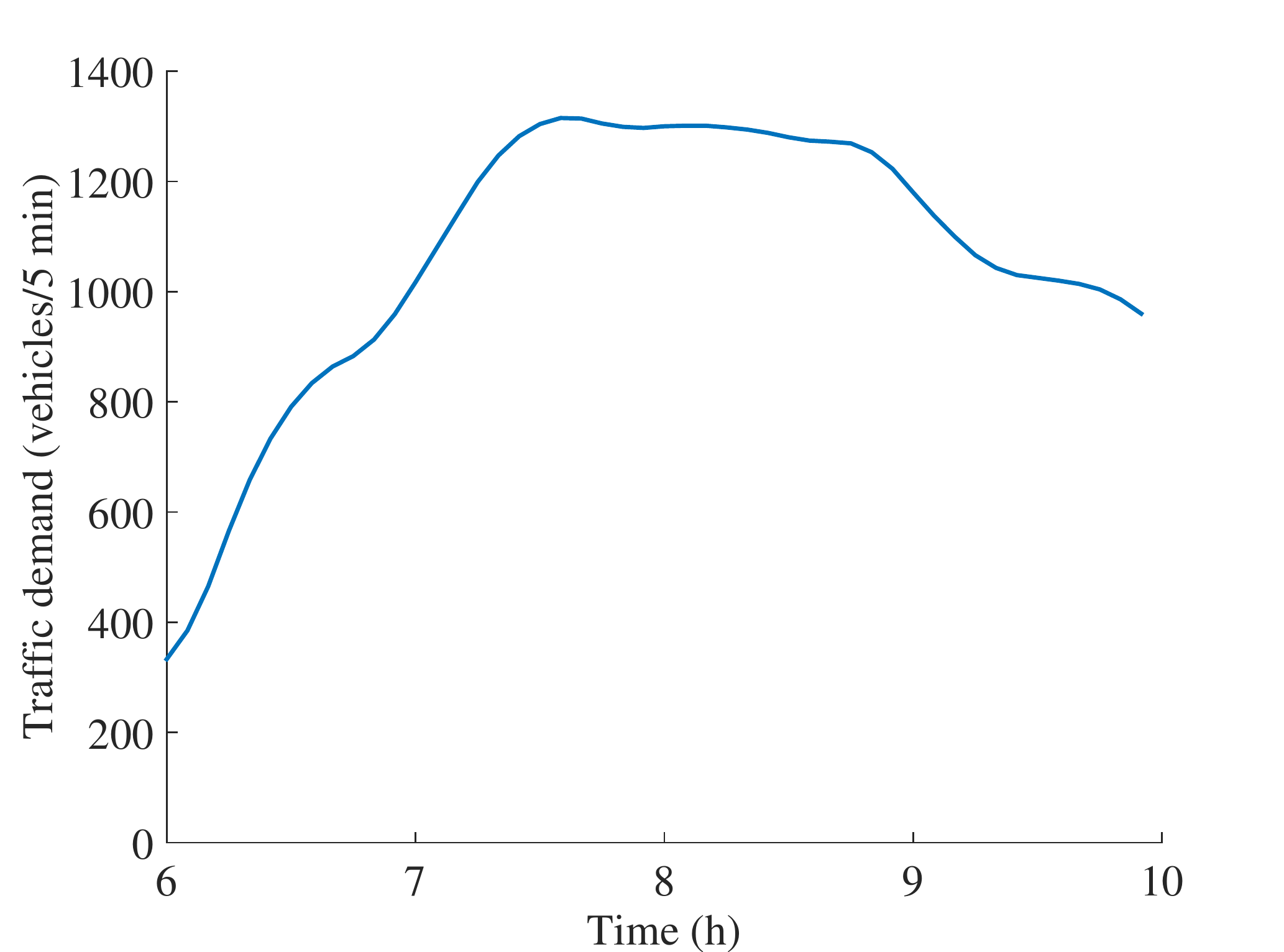}\label{demand}}
\caption{The network of Leidschendam-Voorburg}
\end{figure}

We simulated a part of the municipality of Leidschendam-Voorburg, located in the province of South Holland. The model has been developed and calibrated for another project by the engineering company Sweco. The sequel of this section, firstly presents the basic configuration characteristics  and secondly, the traffic demand  of the simulated network.  The traffic network of the town of Leidschendam-Voorburg includes different types of roads such as freeways, arterial roads and urban streets. It is connected to the A4 highway and the N14 crosses through the city via  tunnels. A map of the area that is simulated is presented in Figure \ref{map}.
The simulated traffic  network of Leidschendam-Voorburg in Paramics covers  an area of about 7 km\textsuperscript{2}  and has a total road length of  29.9 km. It includes 67 zones connected with 491 nodes and 1068 links. The road types that are present in the network are urban roads with speed limits ranging from 30 km/h to 80 km/h and highways with a speed limit ranging from 80 km/h to 120 km/h. In total, the simulated network has 63 junctions, of which 17 are signalized and the rest are flow controlled. The vehicle types in the simulation are single vehicle units, e.g. cars or LGV, bus units, trams and HGV units and bicycles. The network is covered with  65 traffic loop detectors, of which 6 are used for the tram and/ or the bus. The configuration of the simulated network of Leidschendam-Voorburg can be seen in Figure \ref{fig:leidschendam_model}. The location of the loop detectors  across the network's intersections  can be seen in green color.

The simulated time period of the analysis is the morning peak period from 06:00-10:00, when higher congestion is observed in the real network. The profile of the demand during the morning peak period is presented in Figure \ref{demand}. The demand is presented given for 5-minute intervals. As it can be seen, traffic increases gradually with a peak from 07:30 to 09:00 and then, demand decreases until 10:00. 


\begin{figure*}
\subfigure[Ground truth (100\% penetration rate, full data availability)]{\includegraphics[width=\halvepagina]{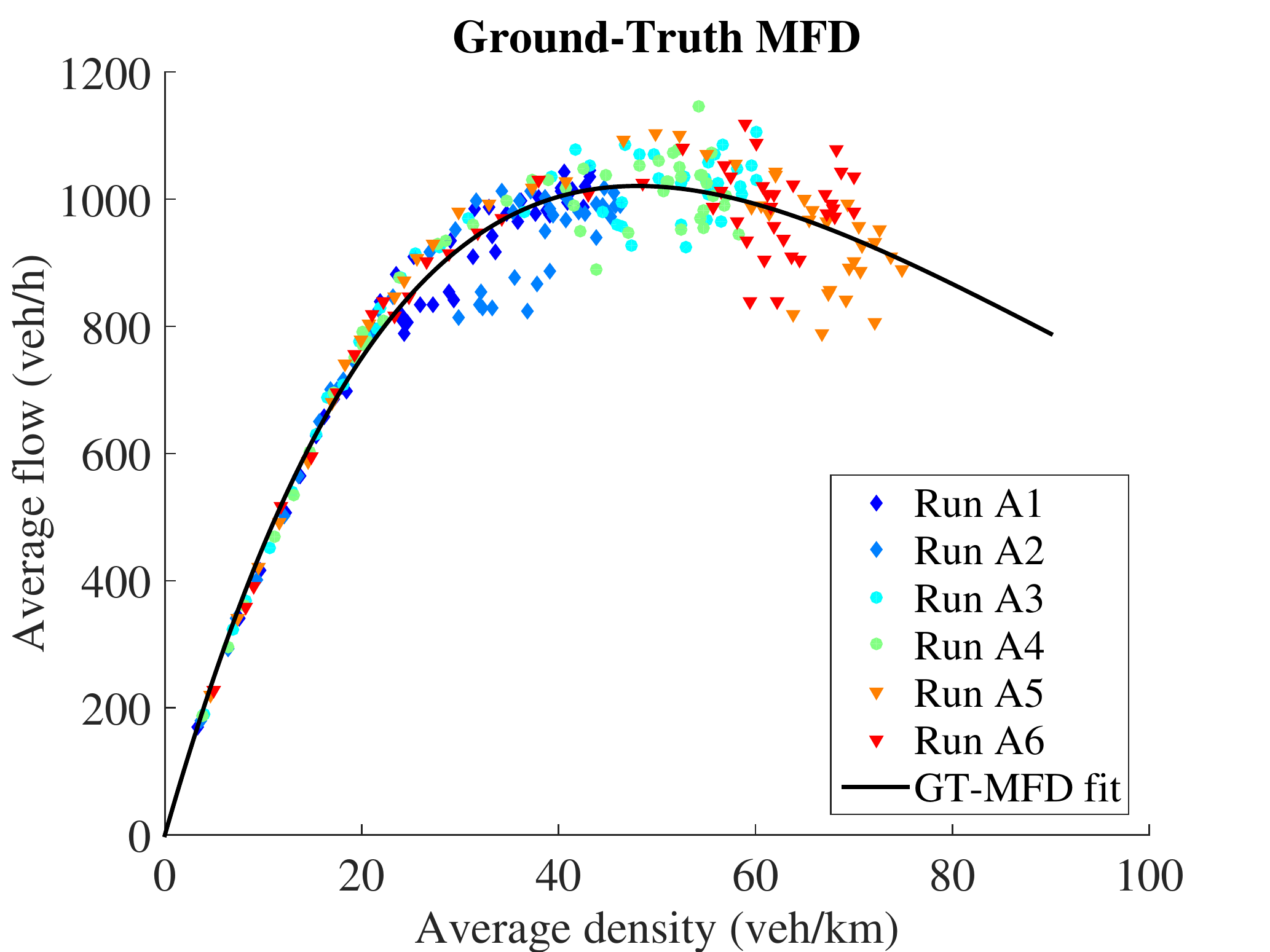}\label{fig:gt_mfd}}
\subfigure[Sampling data and data fusion]{\includegraphics[width=\halvepagina]{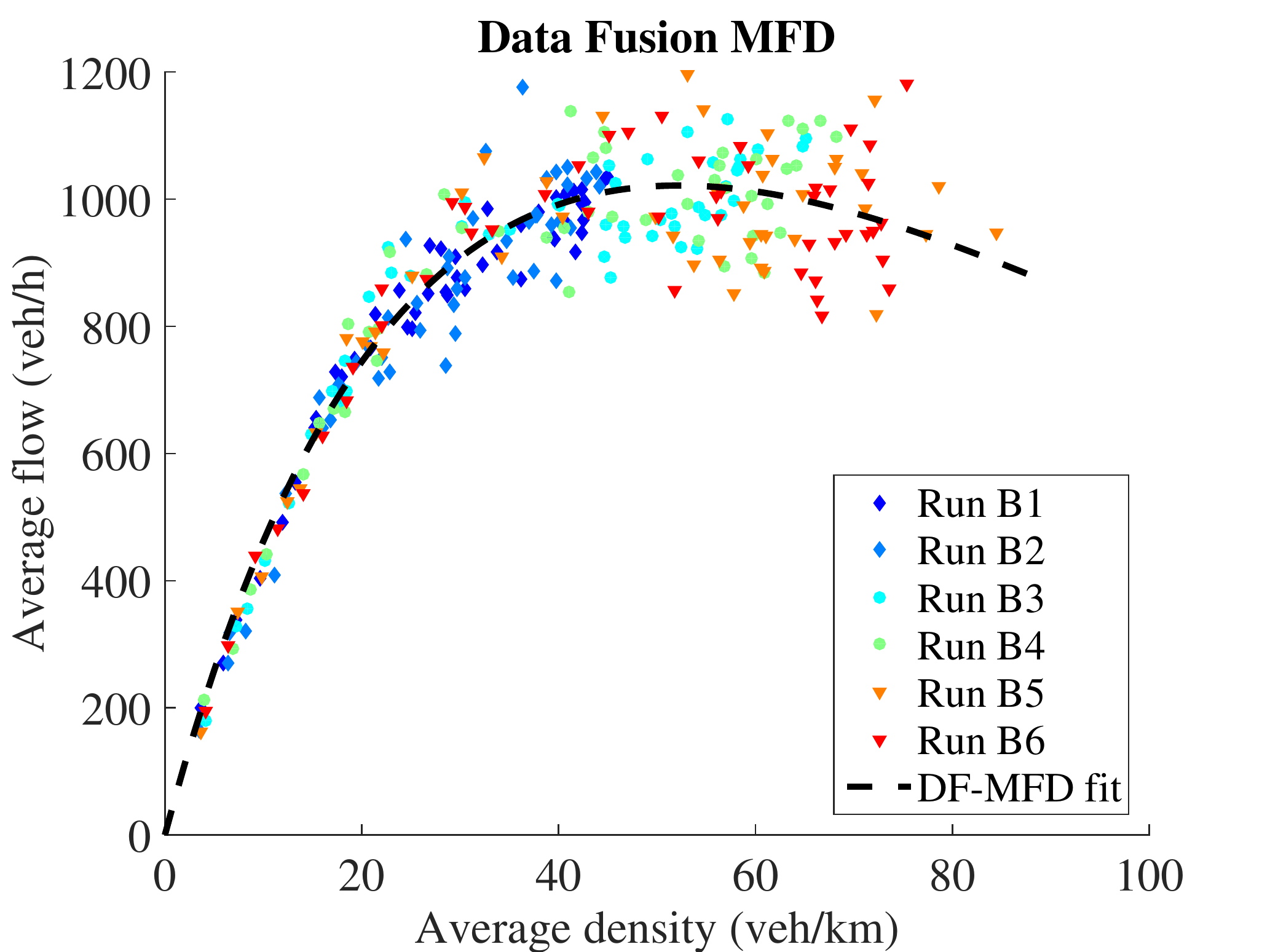}\label{fig:est_mfd}}\\
\subfigure[Comparison of fitted ground truth and data fusion MFD]{\includegraphics[width=\halvepagina]{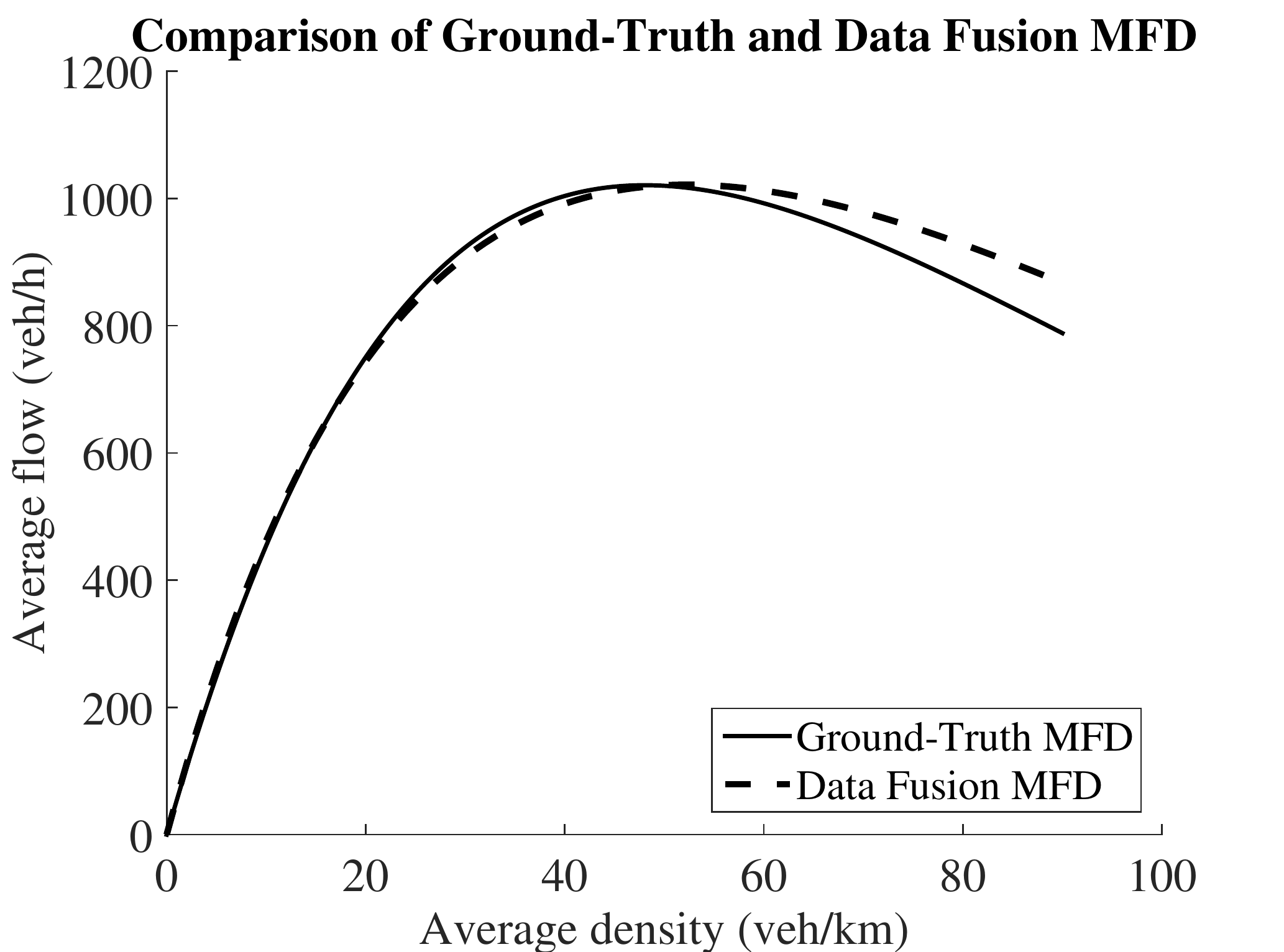}	\label{fig:mfds}}
\subfigure[Comparison between the estimated and the real network densities]{\includegraphics[width=\halvepagina]{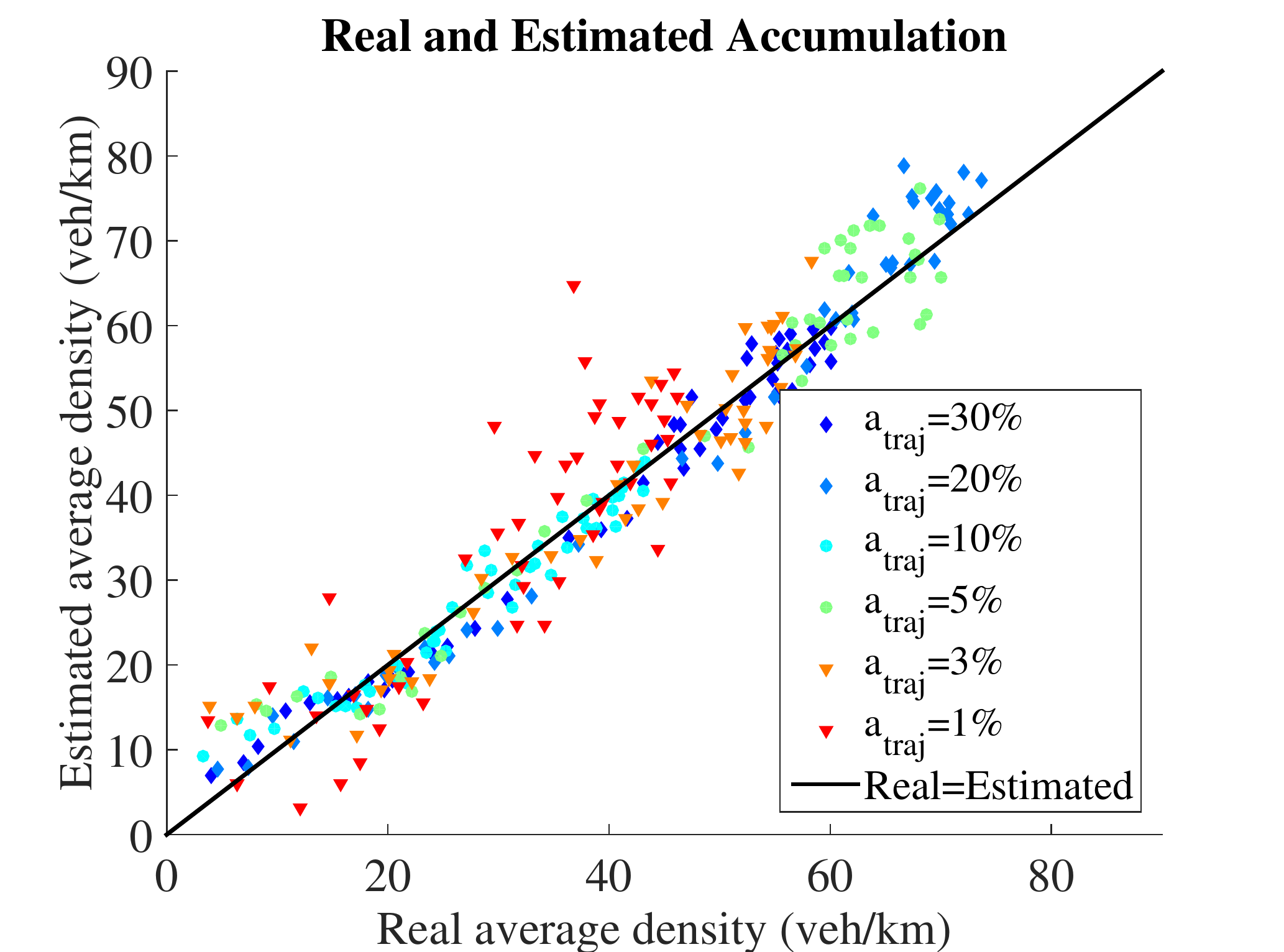}\label{fig:dens_comp}}
	\caption{MFDs: ground truth and data fusion MFD}      
\end{figure*}

\begin{table*}
	\centering
	\caption{Comparison between the parameters of the Data Fusion MFD and the Ground-Truth MFD}
	\label{table_comparison}
	\begin{tabular}{rlcccc}
		\toprule
&&{Properties}& Ground-truth MFD & Data Fusion MFD & Units\\ \midrule
\parbox[t]{2mm}{\multirow{5}{*}{\rotatebox[origin=c]{90}{Parameters \&}}}&
\parbox[t]{2mm}{\multirow{5}{*}{\rotatebox[origin=c]{90}{fitness}}}&
$A$&       -0.02 &        -0.02 &   -     \\
&& $a1$&  54.4       &      58.05  &   -      \\
&& $a2$&    0.19     &      -0.25   &   -     \\
&& Adjusted $R^{2}$ &    0.94     &0.90       & - \\
&& RMSE      &   49.51      & 69.20 & vehicles/h     \\ \midrule 
\parbox[t]{2mm}{\multirow{6}{*}{\rotatebox[origin=c]{90}{Characterising}}}&
\parbox[t]{2mm}{\multirow{6}{*}{\rotatebox[origin=c]{90}{points}}} &&&&\\
&& Free Flow Speed  &     54.4    &    58.05        &  km/h     \\
&&Capacity    &     1021    &1022   &  vehicles/h   \\
&& Critical Density  &      48   &    52  &  vehicles/km\\
&& Critical Speed   &21.3&     19.7 &  km/h   \\&&&&&\\  \midrule 
	\end{tabular}
\end{table*}

In order to get the interval variability in the traffic process, we do 6 different runs. This section will later show that these runs are comparable. 
For each of these runs, we assume a different level of penetration rate. We will check the method for penetration rates of 30\%, 20\%, 10\%, 5\%, 3\% and 1\%. Note that randomly it has been chosen whether the vehicles are sending their speeds, hence they are not necessarily equally distributed over the various links in the network. Besides, different runs also have a different demand level to obtain the all network conditions (from free flowing to congested), for which we take 100\%, 110\% and 120\% of the original demand. The resulting estimated accumulations are compared with the real accumulations taken from 100\% vehicle trajectories to validate their accuracy.

\subsection{Results}\label{sec_simresults} 
This section presents the results of applying the method on the network of Leidschendam-Voorburg. First, the results of the MFD fitting procedure is described, and then the results of step 2, i.e. the traffic state estimation. 

\subsubsection{Constructed MFDs}\label{sec_MFDresults}
First, \fig \ref{fig:gt_mfd} shows the MFD for the case with full data availability (100\% FCD); the shape is crisp, and a clear relationship between average density and average flow is visible. The fit of the MFD follows the points nicely, confirming the functional form chosen. \Fig \ref{fig:est_mfd} shows the MFD based on the data fused from loop detectors and trajectories (limited FDC). The shape is also very crisp, and very similar to the MFD for the full data. Also here, a good fit is possible. To compare the fits, and implicitly see how much the reduction of penetration rate influences the fitted MFD, both fits are plotted in \fig \ref{fig:mfds}. The free flow branch and the peak are almost identical; only for the congested branch, the MFD based on limited data is slightly lower. A comparison of the parameters and the characterising points of both MFDs is found in \tab \ref{table_comparison}. This shows that the parameters and characteristics of the MFDs are very similar. All in all, it seems a MFD can be fitted very well with the available data.  The fit described here is the fit based on a combination of various points (i.e., combining all runs, each having a different penetration, and hence all points in the figure). Next, we will discuss the quality of the fit for the various penetration rates. 

\subsubsection{Quality of traffic states estimation}
Following the process of the second step as described in section \ref{sec_step2}, the traffic state, hence the average density, is estimated. 
\Fig \ref{fig:dens_comp} compares the estimated and real average density. 
The error seems to be rather constant throughout the entire density domain. 
\begin{figure}
\includegraphics[width=\halvepagina]{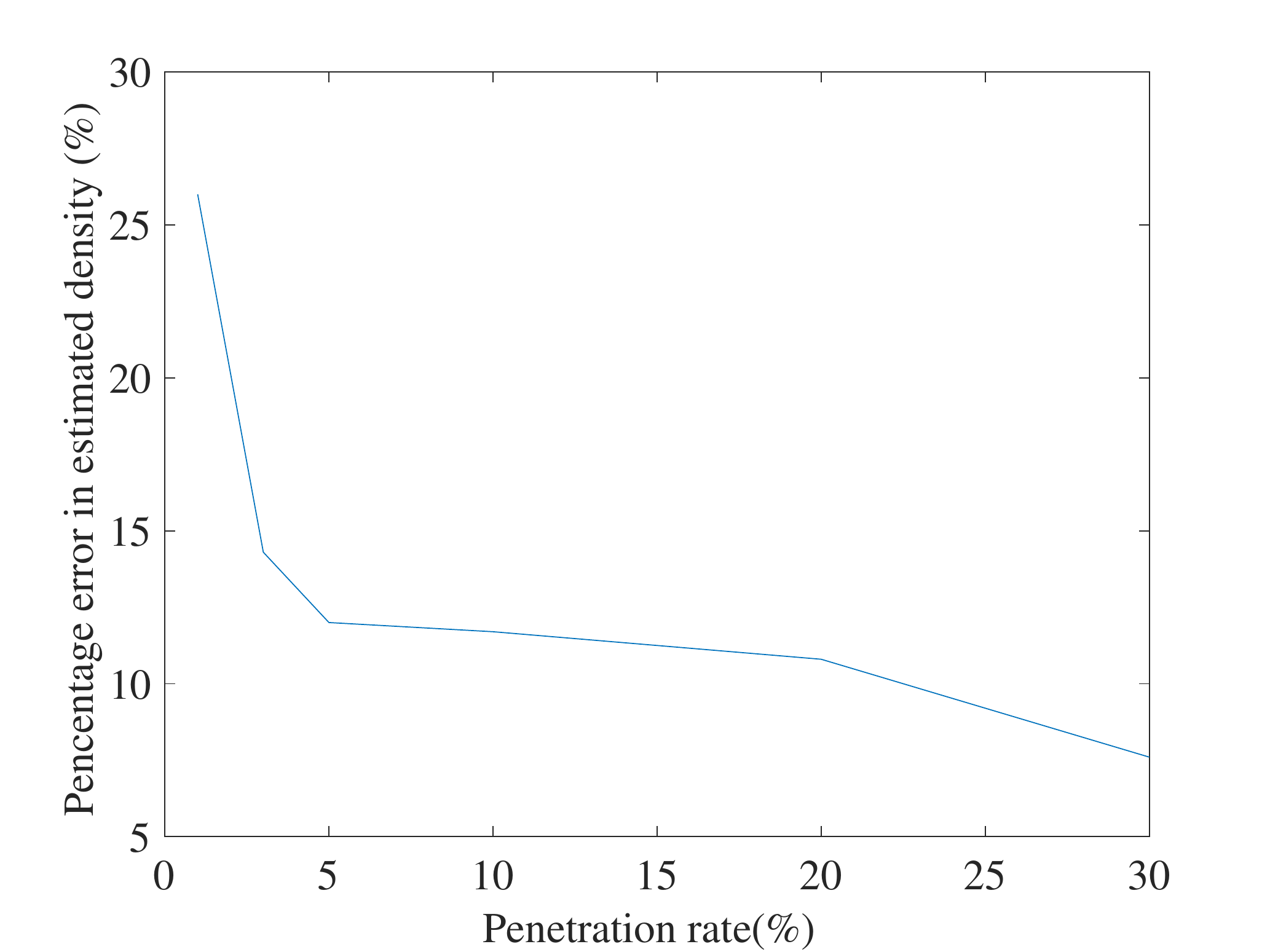}
\caption{Mean percentage error for the densities estimated at various penetration rates $\penrate$}\label{dens_comp_figure}
\end{figure}
\Fig \ref{dens_comp_figure} shows the mean percentage error for each fraction of vehicle trajectories. It shows that the higher the fraction of the known vehicle trajectories is, the lower the mean error. This means that the estimated densities are more similar to the reality when the penetration rate increases, which is in line with the expectations.  For the lowest penetration rate tested (1\%), the error is 26\%. The error almost halves to 14\% with 3\% penetration rate. Then it remains relatively constant up to a penetration rate of 20\% . For the highest penetration rate (30\%), the error is 7.6\%. Note that these reported errors are the errors after the fitted MFD is applied. Hence, for a 100\% penetration rate, these errors do not go to 0, but the remaining error shows the natural stochastic fluctuations of traffic. There are fluctuations for the fitness for varying draws of the penetration rate. These fluctuations are small, ranging from 0.52 veh/km (standard deviation) for 1\% penetration rate decreasing to 0.15 veh/km for 30\% penetration rate.

These results can offer a first glance at how the different penetration rates influence the accuracy of the estimation of the traffic state. Nevertheless, these results should only be considered as  validation test on how well the proposed process works under different conditions. If it is desired to extract conclusions on the required penetration rate of floating car data, additional information is needed  on whether an exact traffic state estimation is desired or it is sufficient to only know whether the network is in the congested state or not. The desired accuracy level can be determined depending on the purpose that the traffic state estimation will be used for.

\section{Reliability of estimate}\label{sec_reliability}

Whereas the previous section has shown an estimation method for the traffic state, this section will present the reliability of the method. This is first mathematically derived, and then the confidence interval will be shown for the example case as described above. Section \ref{sec_incident} will test a measure to identify whether the estimated traffic state can be considered unreliable, resulting from an abnormal event within the network. 

\subsection{Propagation of errors}\label{sec_uncertainty}
\begin{figure*}[t!]
\subfigure[MFD with confidence intervals]{\includegraphics[width=\halvepagina]{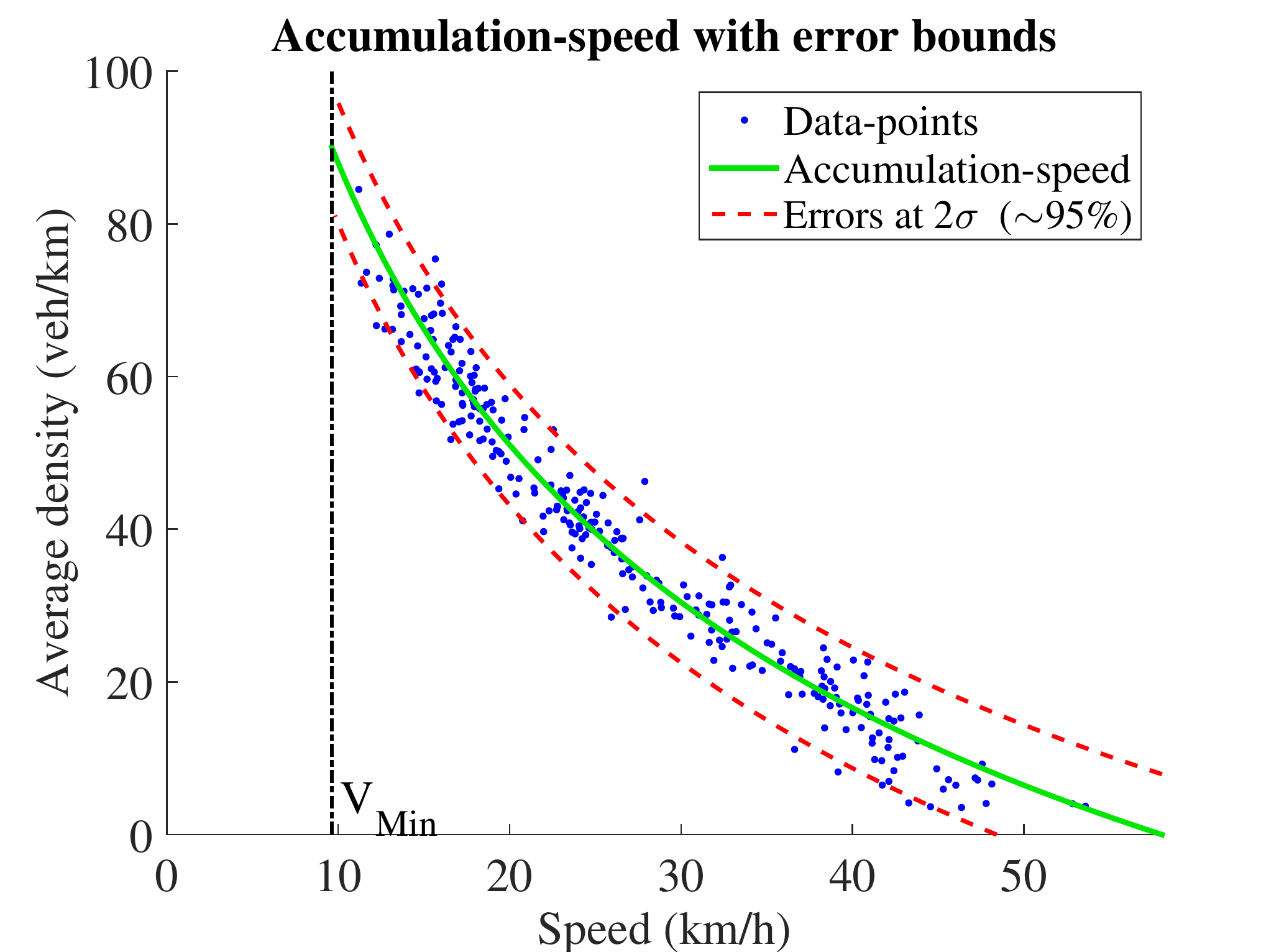}\label{fig_kv_MFD_bounds}}
\subfigure[Schematic representation of the combination of the probabilities]{\includegraphics[width=\halvepagina]{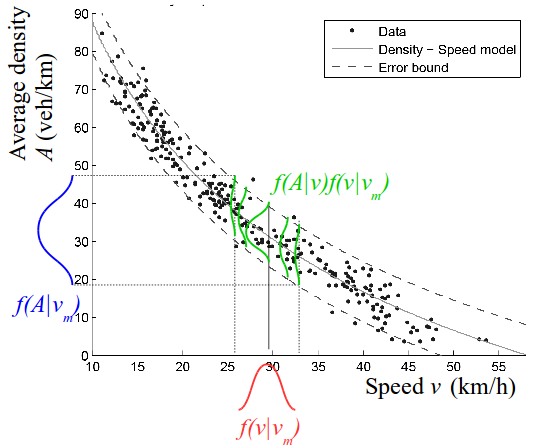}
\label{fig_conceptuncertenty}}\\
\subfigure[Probability of the estimated density values  given speed]{\includegraphics[width=\halvepagina]{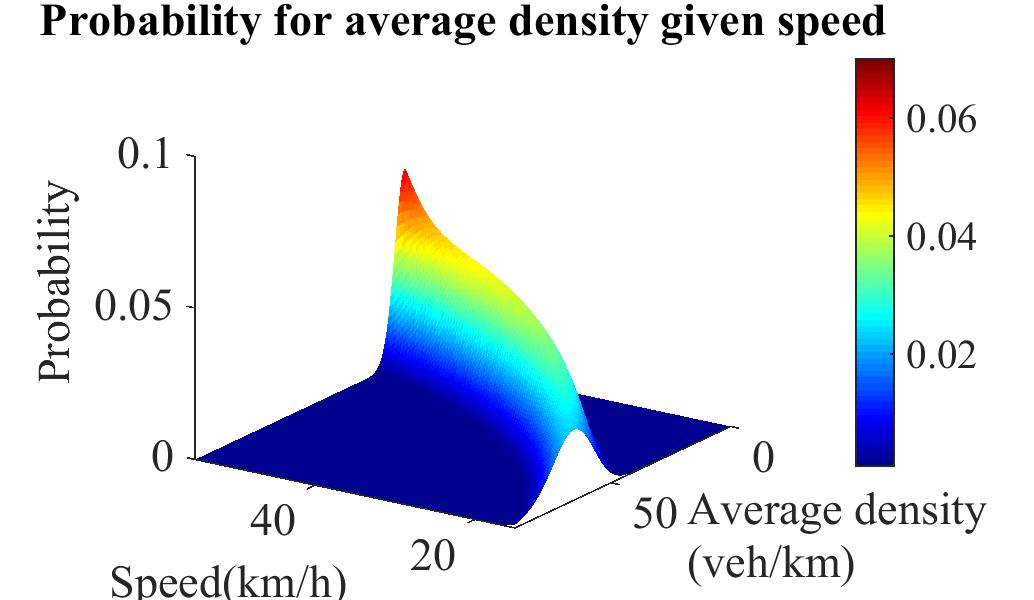}\label{fig_3duncertenty}}
\subfigure[Probabilities of estimated densities given the speed measurements]{\includegraphics[width=\halvepagina]{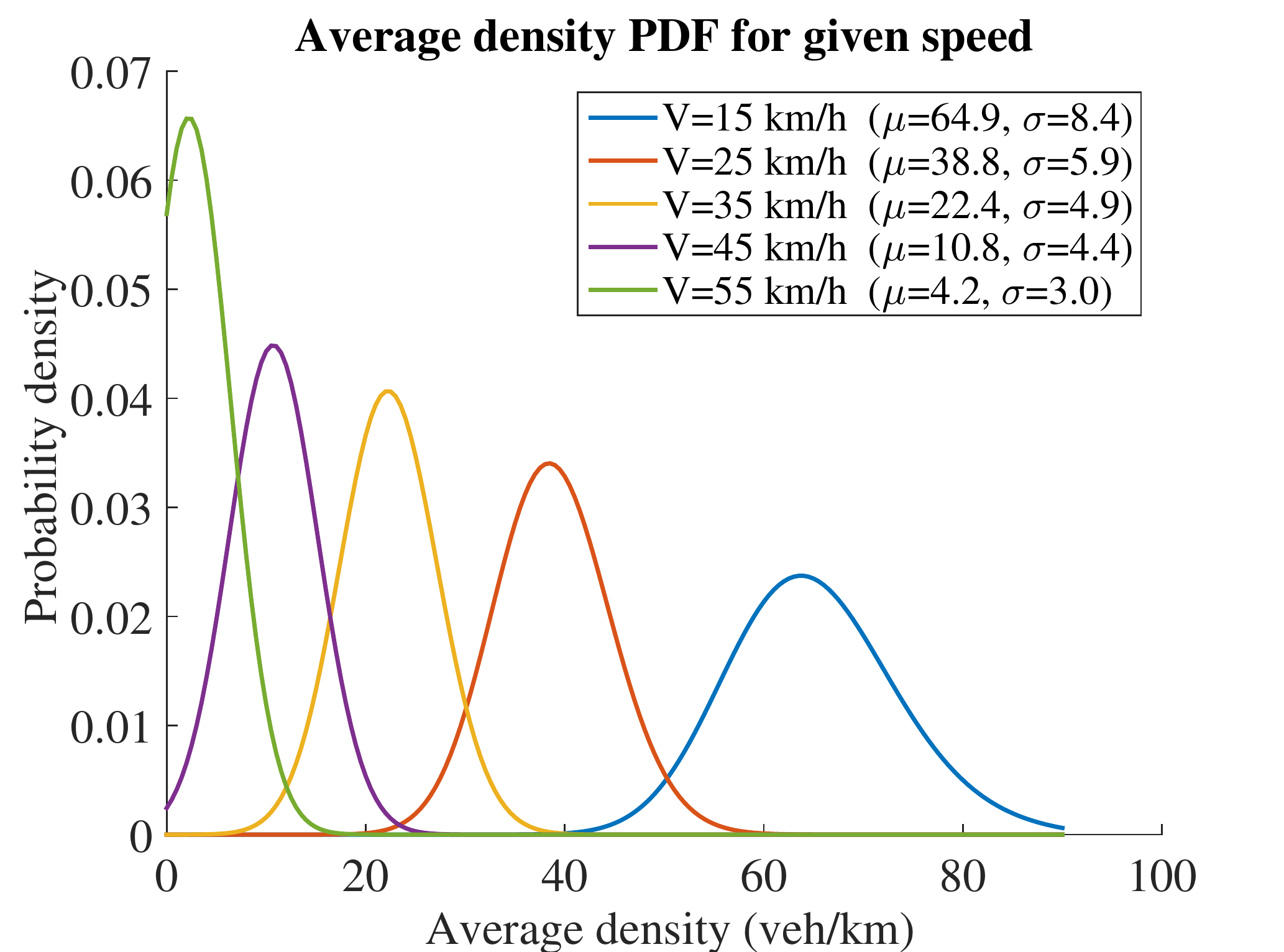}\label{fig_examples_speeddensity}}
\caption{Combined errors in estimation}
\end{figure*}

Both steps of the process of section \ref{sec_method} (i.e., finding the shape of the MFD and getting the state from the speed of the FCD)  are subject to an error. To illustrate the first error, in the fitting of the MFD, error bounds are added in the speed-average density plane, see \fig \ref{fig_kv_MFD_bounds}. The bounds  seem to widen for higher densities, but that is an optical illusion. The width of the bounds is twice the standard deviation of the points to the line, for both directions. 
If the errors between the points and the line are normally distributed, 95\% of the points fall within the bounds. 
For each real speed, there is a probability distribution of the matching density, which we indicate by $f(\accumulation|\vreal)$.  

In the second step, an error is introduced because the speeds of the sample of vehicles might differ from the speed of all vehicles. We relate the estimated speed to the real speed -- using the ground truth information from the simulation, and similar to the densities, we determine the confidence intervals. 
We denote the probability density function that speed has a value $\vreal$ under the mean of the sampled speed $\vest$ as $f(\vreal|\spdm)$.

Both errors have a joint effect, which is discussed in the sequel of this paragraph. Given a speed measurement $\spdm$, we want to estimate the probability density function $f(\accumulation|\spdm)$ of the error level of an average density occurring given this speed measurement. What we have is the probability density function of the average density given the real speed measurements  $f(\accumulation|\vreal)$ and the probability density function of the measured speeds  given the real speeds $f(\spdm|\vreal)$. In this section we combine the uncertainties, as graphically indicated in \fig \ref{fig_conceptuncertenty}. Basically, all joint elements of ($\accumulation, \spdm$) are given a weight, such that a cross section of the graph in \fig \ref{fig_conceptuncertenty} (later constructed in \fig \ref{fig_3duncertenty}) for a constant $\spdm$ gives a function for the average density $\accumulation$. Scaling this function with a scalar such that the integral of that function over density is one, gives the probability density function. We will now present this system in equations.

We consider the average density $\accumulation$ and the speed values $\spdm$ as dependent events, so  the conditional probability of $\accumulation$ given $\spdm$ is defined as (Kolmogorov definition):
\begin{equation} \label{cond_prob}
f(\accumulation|\spdm)=\frac{f(\accumulation,\spdm)}{f(\spdm)}
\end{equation}
where $f(\accumulation,\spdm)$ is the joint probability density function of aveage density and speed. 
Using conditional probabilities, we can rewrite:
\begin{equation}
f(\accumulation,\spdm)=\int_{0}^{\vmax}f(\accumulation|\spdm,\vreal)f(\spdm|\vreal) f(\vreal) d\vreal
\end{equation}
In this equation $\vreal$ is given. If one knows the real speed, one does not need to estimate so, and this simplifies to 
\begin{equation} 
f(\accumulation,\spd)=
\int_{0}^{\vmax}
f(\accumulation|\vreal) f(\spdm|\vreal) f(\vreal) d\vreal
\label{eq_fkv}
\end{equation}
Similarly, using conditional probabilities we can rewrite $f(\spdm)$ to
\begin{equation}
f(\spdm)=\int_{0}^{\vmax}f(\spdm|\vreal) f(\vreal) d\vreal \label{eq_fv}
\end{equation}
Substituting \eq \ref{eq_fkv} and \eq \ref{eq_fv} into \eq \ref{cond_prob}, we obtain after some algebraic manipulation:
\begin{multline}
f(\accumulation|\spdm)=\\
\frac{\int_{0}^{\vmax}f(\accumulation|\vreal) f(\spdm|\vreal)f(\vreal) d\vreal}{\int_{0}^{\vmax}f(\spdm|\vreal) f(\vreal) d\vreal}=\\
\frac{\int_{0}^{\vmax}f(\accumulation|\vreal) f(\spdm|\vreal) d\vreal}{\int_{0}^{\vmax}f(\spdm|\vreal)  d\vreal}
\end{multline}
Assuming a maximum average speed $\vmax$ of 60 km/h (a high estimate under urban conditions), we know all right-hand side elements of the equation and we can numerically compute the conditional probability.

The resulting probability density function is shown in \fig \ref{fig_3duncertenty}. \Fig \ref{fig_examples_speeddensity} shows some examples of resulting curves for various measured speed $\spdm$.  Higher speeds lead to (absolute) higher uncertainty for higher average densities, but the uncertainty relative to the average density decreases with higher average densities. 

\subsection{Effect of exceptional conditions}\label{sec_incident}
In abnormal conditions, the MFD might not hold, which voids the method proposed here. An important question in this respect is to which extent we can recognize these situations from the data. We are hence looking for an indicator predicting the reliability of the estimate. We propose to consider the \emph{rate of change of the speed as function of time}. An incident would cause cars to stop upstream of the incident, without cars moving downstream of the incident. Therefore, network speed would reduce more rapidly than usual because the jam will grow rapidly since the outflow is lower than the regular capacity.

To test the working of the method under incidents, as well as the working of the above-mentioned indicator, we simulate different scenarios with incidents. Various scenarios are tried, each with a different location of the incident. The  locations of the incidents in the network can be seen in Figure \ref{incident_locations}. Some incident locations are already highly congested (intersection with the four incidents A, C, D and E),  one is on the highway (G),  one in the network center (B) and one on a smaller road of less significance (F). The road incidents are simulated for one hour from 07:00-08:00. During the road incidents, one lane is blocked, reducing the capacity of the respective road segment.
\begin{figure*}[t!]
\subfigure[Estimated MFDs with incidents]{\includegraphics[width=\halvepagina]{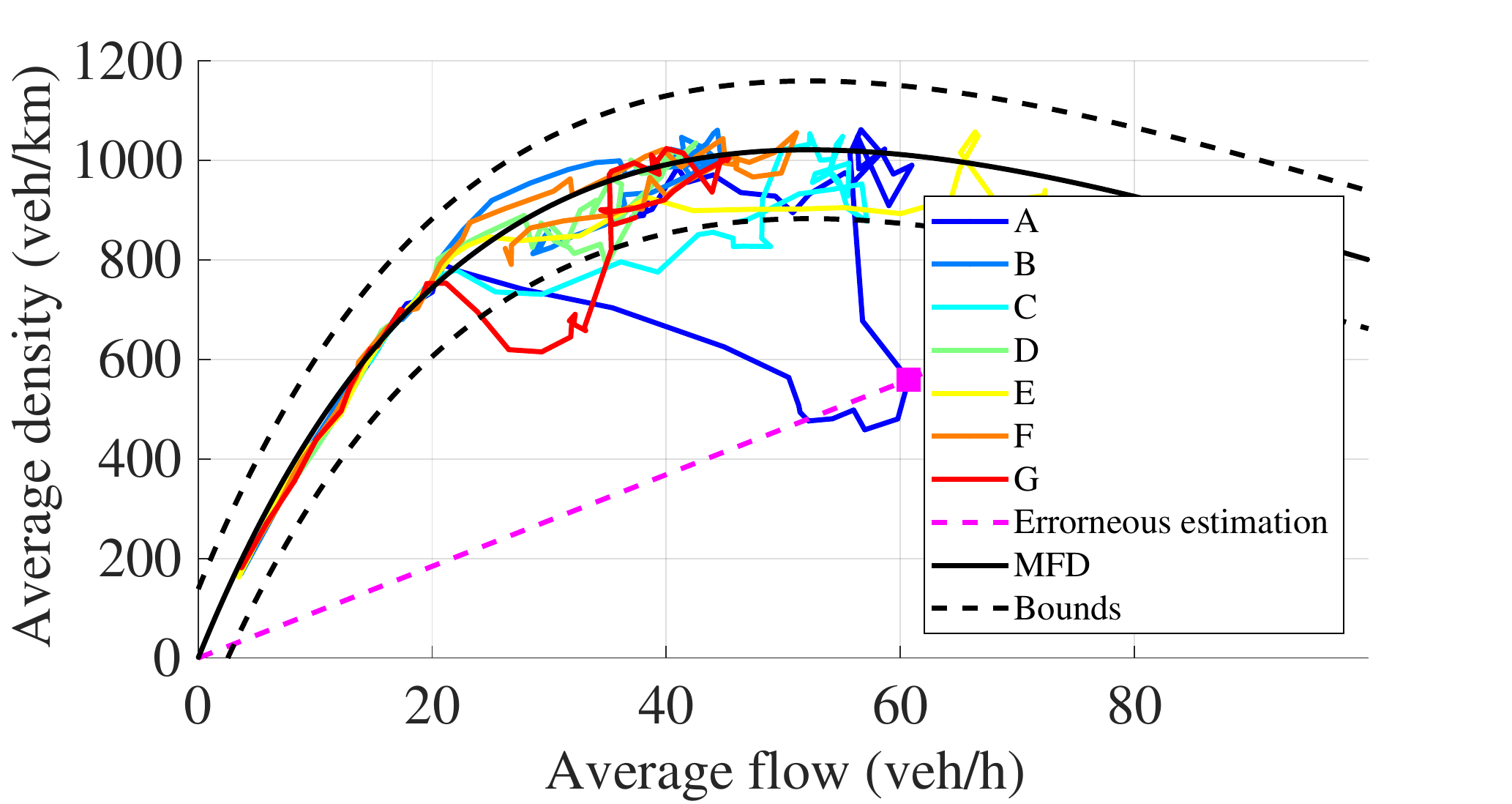}\label{incident_mfd}}
\subfigure[Speed measurements during the simulation of the incidents]{\includegraphics[width=\halvepagina]{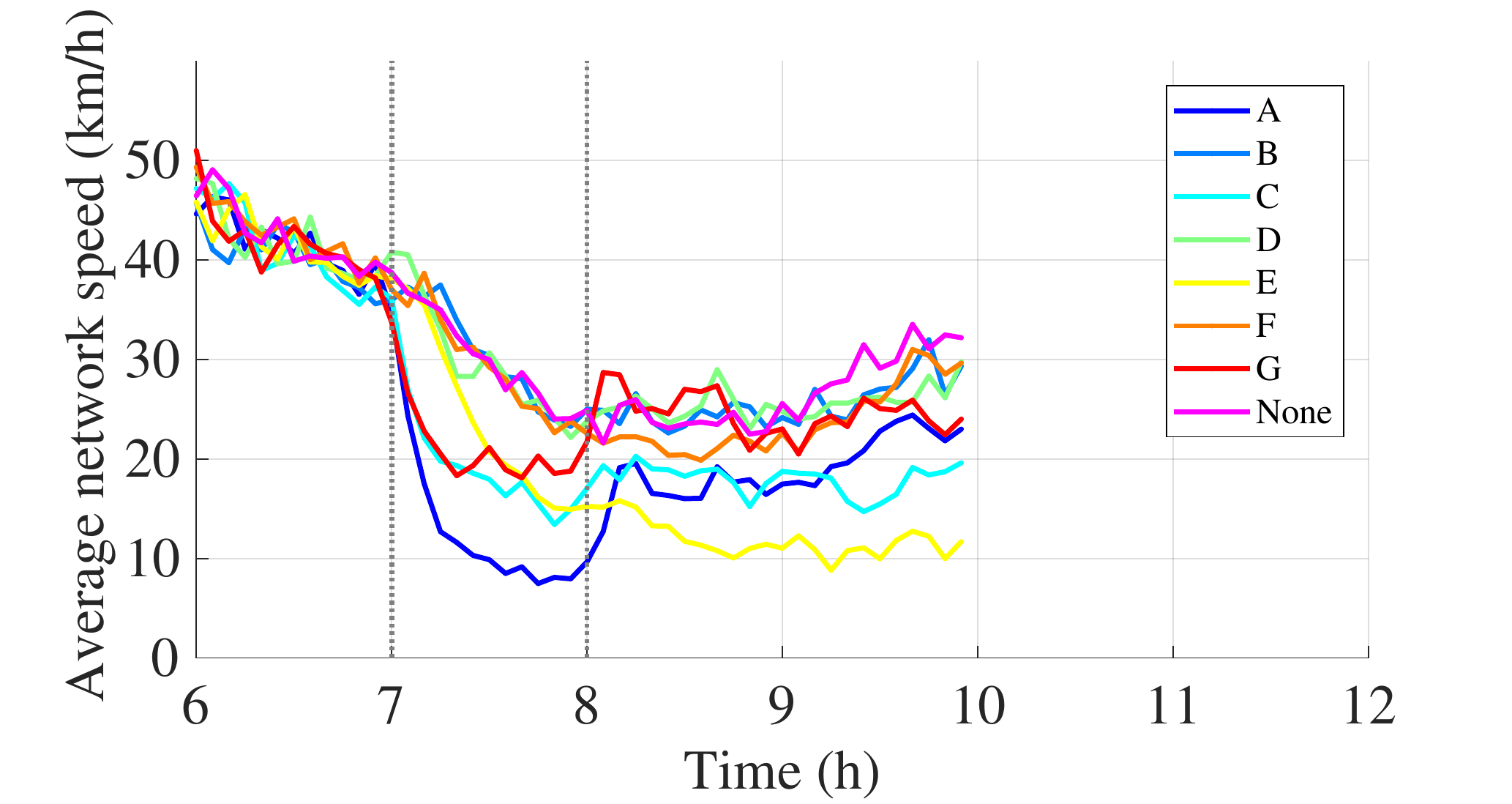}\label{incident_speed}}\\
\subfigure[Speed differences after 5 minutes]{\includegraphics[width=\halvepagina]{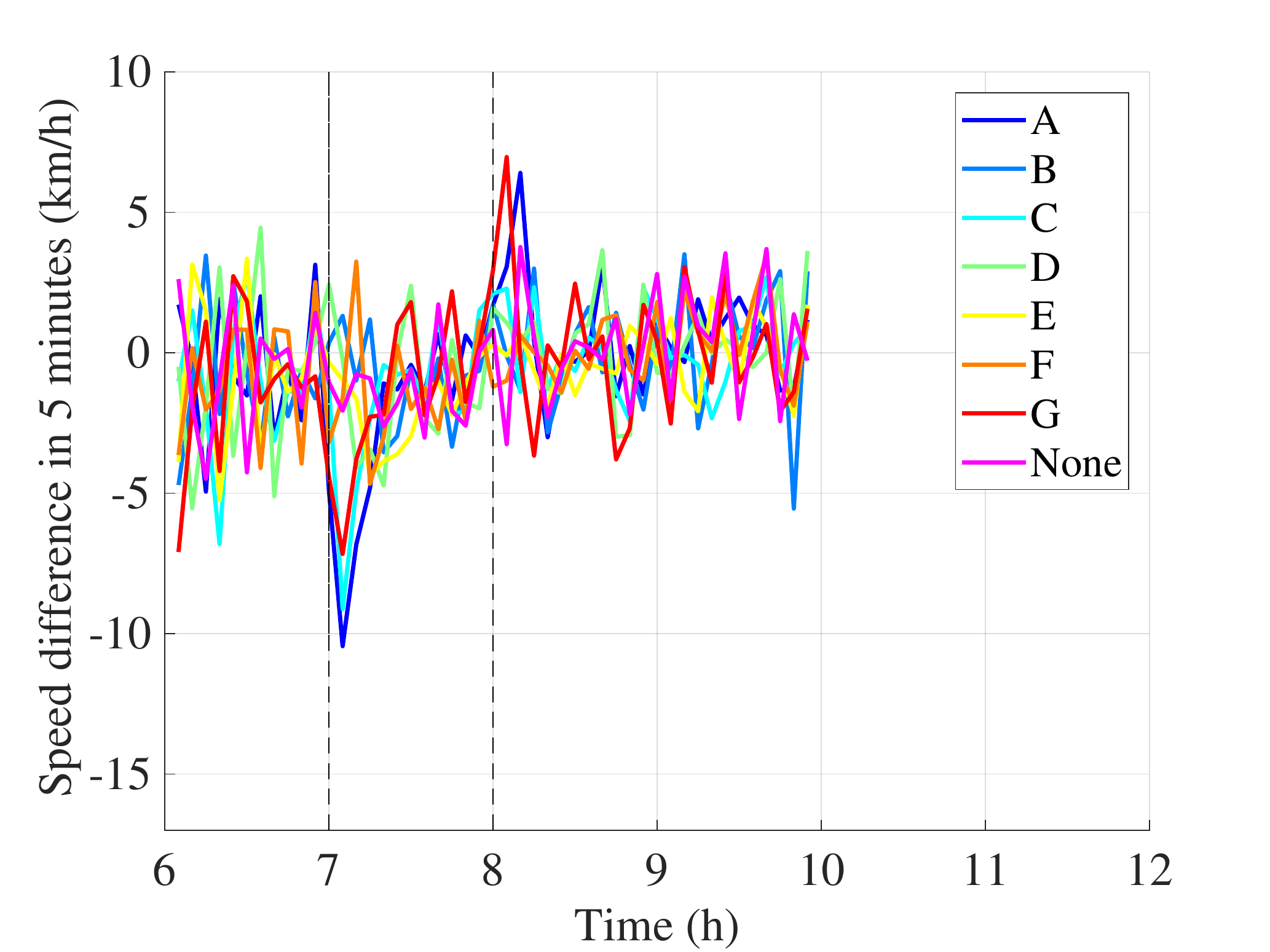}\label{speeddiff5mins}}
\subfigure[Speed differences after 10 minutes]{\includegraphics[width=\halvepagina]{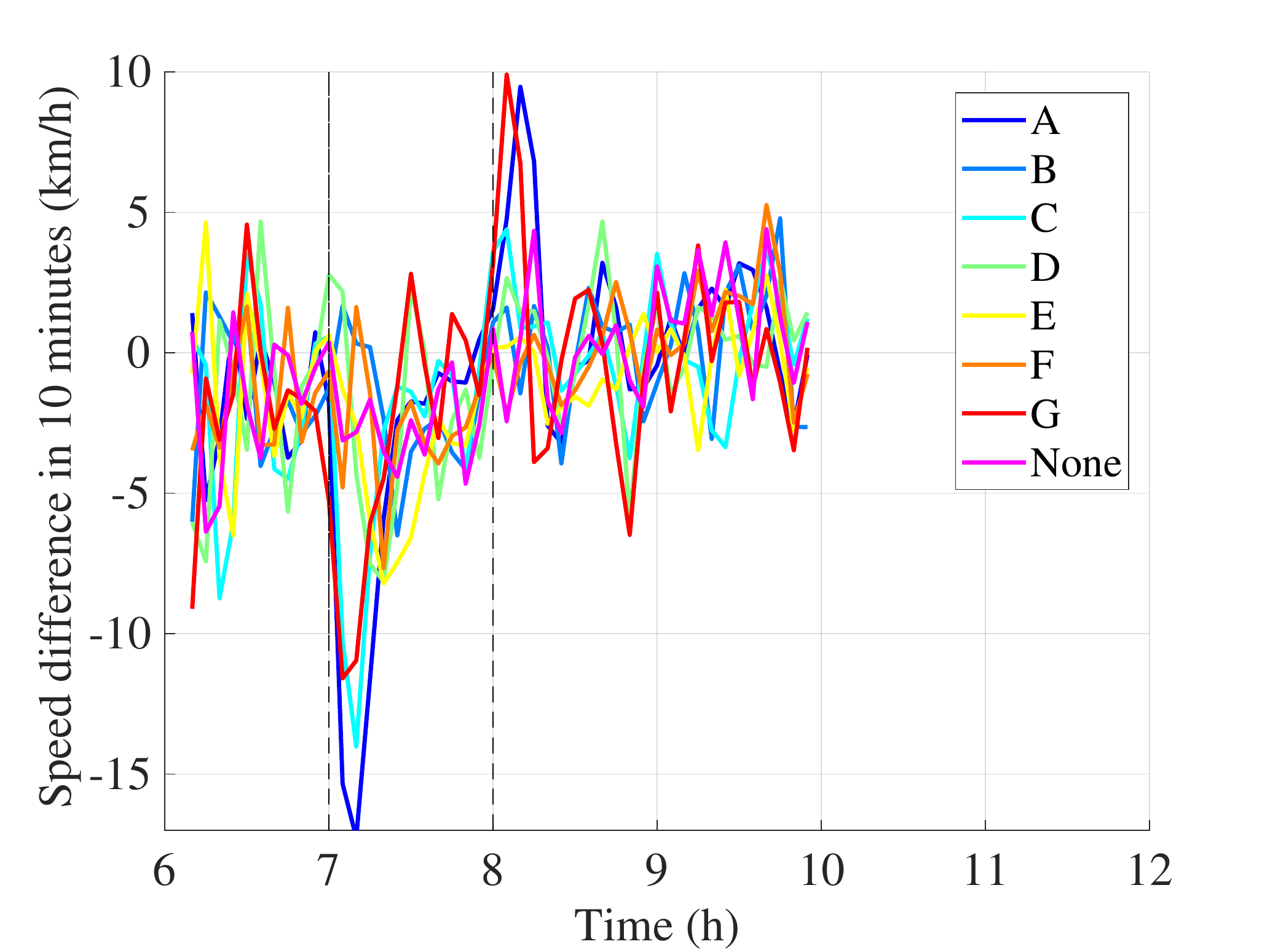}\label{speeddiff10mins}}
\caption{Effect of the incidents. Different colored lines are different incident scenarios. The incident was simulated from 7 to 8, indicated with dashed black lines (fig b-d).}
\end{figure*}

Depending on the location, the traffic state differs. \Fig \ref{incident_mfd} shows the MFDs created with sample data. The speeds at incident locations B, D, and F fall within the bounds.  For other scenarios the speed falls outside the bounds: scenarios with incidents at location A, C, E (slightly), and G. In all these cases, the speed is lower than the speed without incident. That means that the indicator should predict that the state estimated during these incidents is not reliable. 

Let's consider what would happen if there was no reliability indicator. For instance the traffic state of the magenta square at the blue line in \fig \ref{incident_mfd}. The estimated traffic state would be the point where a line with the estimated speed (magenta dotted line) meets the MFD, i.e. at an average density of slightly over 90 veh/km. In reality, the average density is around 63 veh/km. This erroneous estimation happens for the lines which are out of the confidence interval. 

We will now show an indicator, which will show whether the estimated traffic state is out-of-the-ordinary, and therefore the traffic state will lie outside the band, and the estimation procedure is not working properly. Since speed is the basis for our method, we will also use speed as basis for the reliability indicator. 

As can be seen in \fig \ref{incident_speed}, speeds drop rapidly if there are incidents for which the MFD falls outside the uncertainty bands. \Fig \ref{speeddiff5mins} and \ref{speeddiff10mins} show the speed differences over a 5 and 10 minute interval respectively. It shows that for incidents at spots for which the MFD falls outside the bounds, also show the largest speed drop. A threshold for the speed drop can be an indicator to which extent the network operates at its normal operations, and hence whether the method developed in this paper can be used to estimate the traffic state.

The value for this speed threshold is expected to differ per network, both in time as well as in magnitude of the speed drop. A larger time over which a drop should be observed could give a higher reliability for the reliability indicator. However, that would also delay the signalling of incidents. A combination of times, one for early signalling, and one for more reliable signalling, could also be used.   A larger network is expected to suffer less from an incident, simply because the affected part is a smaller fraction of the network. At the other hand, also the bounds are probably tighter due to less effect of stochastic fluctuations in normal conditions. Ultimately, the testing on the network itself should show the network-specific bounds. 

For our network, a speed reduction of 5.75 km/h in 5 minutes was a good threshold. For incident sites B, D, and F, reductions never exceeded this threshold. For incident sites C, G, and especially A, the change well exceeded the threshold. For site E, the change is around the threshold. This is very well aligned with the bounds on the MFD. Incidents for which the change exceeds the threshold fall outside the bounds, and vice versa. Similar, and more reliably, a threshold for a speed reduction of 8.5 km/h in  a 10-minute interval coincides well with falling inside or outside the reliability bounds. 

Thus, we find that the estimation procedure and the bounds are useful also  in case of small network disruptions. The evolution of estimated network speeds over time is a good indicator for the size of the network disruption. If it does exceed a threshold value, the method should not be used. 

\section{Conclusions}\label{sec_conclusions}

In this paper we developed a two-step estimation method for average network density by fusing loop detector data and low penetration rates of probe vehicle data. In the first step, a calibration step, we use loop detector data and floating car data (of a limited number of vehicles) to construct the MFD. In the second step, the traffic state (network density) is estimated based on the speed of a limited number of probe vehicles and the MFD. We explicitly address the uncertainty in the process, caused by the creation of the MFD based on data fusion, as well as the uncertainty in the traffic speed given the speed of the FCD.

We conclude the following:
\begin{itemize}
	\item The main conclusion is that there is a simple, easily implementable method which works at acceptable accuracy at low penetration rates. The MFD provided for the operational speed estimate can be initialized with loop detector data and floating car data. Through simulation we show a mean estimation error of less than 15\% for penetration rates of 3\% and higher. 
	\item The method also endogenously provides an error bound to the estimate of average density. The accuracy of estimation is in absolute terms better for lower average densities; the relative accuracy is better for higher average densities. In both cases, the resulting traffic estimates are well differentiable (i.e., one knows the approximate level of service for a network).
	\item Finally, a method has been presented which can predict the reliability of the method using the FCD. This is based on the rate of change of the speed as function of time. As long as that is not exceeding a threshold, the method can be used. If the speed changes too rapidly, this indicates that a major disruption has happened and the method should not be used.
\end{itemize}

The method has been tested for a simulated network. Indeed, traffic dynamics for a real network might differ. The goal of this paper is to check the consistency between the estimation method and the ground truth, which is remarkably good. Next steps are to test the method on a real network, using real data.

\bibliographystyle{trb}
\bibliography{state_estimation,references,20170120}

\end{document}